\documentclass[12pt]{article}
\usepackage{graphicx}
\usepackage{lscape}
\usepackage{citesort}
\usepackage{amssymb}
\usepackage{appendix}
\usepackage{multirow}

\begin{document}
\def\qq{\langle \bar q q \rangle}
\def\uu{\langle \bar u u \rangle}
\def\dd{\langle \bar d d \rangle}
\def\sp{\langle \bar s s \rangle}
\def\GG{\langle g_s^2 G^2 \rangle}
\def\Tr{\mbox{Tr}}
\def\figt#1#2#3{
        \begin{figure}
        $\left. \right.$
        \vspace*{-2cm}
        \begin{center}
        \includegraphics[width=10cm]{#1}
        \end{center}
        \vspace*{-0.2cm}
        \caption{#3}
        \label{#2}
        \end{figure}
	}
	
\def\figb#1#2#3{
        \begin{figure}
        $\left. \right.$
        \vspace*{-1cm}
        \begin{center}
        \includegraphics[width=10cm]{#1}
        \end{center}
        \vspace*{-0.2cm}
        \caption{#3}
        \label{#2}
        \end{figure}
                }

\def\ds{\displaystyle}
\def\beq{\begin{equation}}
\def\eeq{\end{equation}}
\def\bea{\begin{eqnarray}}
\def\eea{\end{eqnarray}}
\def\beeq{\begin{eqnarray}}
\def\eeeq{\end{eqnarray}}
\def\ve{\vert}
\def\vel{\left|}
\def\ver{\right|}
\def\nnb{\nonumber}
\def\ga{\left(}
\def\dr{\right)}
\def\aga{\left\{}
\def\adr{\right\}}
\def\lla{\left<}
\def\rra{\right>}
\def\rar{\rightarrow}
\def\lrar{\leftrightarrow}  
\def\nnb{\nonumber}
\def\la{\langle}
\def\ra{\rangle}
\def\ba{\begin{array}}
\def\ea{\end{array}}
\def\tr{\mbox{Tr}}
\def\ssp{{\Sigma^{*+}}}
\def\sso{{\Sigma^{*0}}}
\def\ssm{{\Sigma^{*-}}}
\def\xis0{{\Xi^{*0}}}
\def\xism{{\Xi^{*-}}}
\def\qs{\la \bar s s \ra}
\def\qu{\la \bar u u \ra}
\def\qd{\la \bar d d \ra}
\def\qq{\la \bar q q \ra}
\def\gGgG{\la g^2 G^2 \ra}
\def\q{\gamma_5 \not\!q}
\def\x{\gamma_5 \not\!x}
\def\g5{\gamma_5}
\def\sb{S_Q^{cf}}
\def\sd{S_d^{be}}
\def\su{S_u^{ad}}
\def\sbp{{S}_Q^{'cf}}
\def\sdp{{S}_d^{'be}}
\def\sup{{S}_u^{'ad}}
\def\ssp{{S}_s^{'??}}

\def\sig{\sigma_{\mu \nu} \gamma_5 p^\mu q^\nu}
\def\fo{f_0(\frac{s_0}{M^2})}
\def\ffi{f_1(\frac{s_0}{M^2})}
\def\fii{f_2(\frac{s_0}{M^2})}
\def\O{{\cal O}}
\def\sl{{\Sigma^0 \Lambda}}
\def\es{\!\!\! &=& \!\!\!}
\def\ap{\!\!\! &\approx& \!\!\!}
\def\ar{&+& \!\!\!}
\def\ek{&-& \!\!\!}
\def\kek{\!\!\!&-& \!\!\!}
\def\cp{&\times& \!\!\!}
\def\se{\!\!\! &\simeq& \!\!\!}
\def\eqv{&\equiv& \!\!\!}
\def\kpm{&\pm& \!\!\!}
\def\kmp{&\mp& \!\!\!}
\def\mcdot{\!\cdot\!}
\def\erar{&\rightarrow&}


\def\simlt{\stackrel{<}{{}_\sim}}
\def\simgt{\stackrel{>}{{}_\sim}}


\title{
         {\Large
                 {\bf
Determination of the $\Sigma$--$\Lambda$ mixing angle from  QCD sum rules
                 }
         }
      }

\author{\vspace{1cm}\\
{\small T. M. Aliev \thanks {e-mail:
taliev@metu.edu.tr}~\footnote{permanent address:Institute of
Physics,Baku,Azerbaijan}\,\,, T. Barakat \thanks {e-mail:
tbarakat@KSU.EDU.SA}\,\,, M. Savc{\i} \thanks
{e-mail: savci@metu.edu.tr}} \\
{\small Physics Department, Middle East Technical University,
06800 Ankara, Turkey }\\
{\small $^\ddag$ Physics and Astronomy Department, King Saud University, Saudi Arabia}}

\date{}

\begin{titlepage}
\maketitle
\thispagestyle{empty}

\begin{abstract}

The $\Sigma$--$\Lambda$ mixing angle is calculated in framework of the QCD sum
rules. We find that our prediction for the mixing angle is $(1.15\pm 0.05)^0$
which is in good agreement with the quark model prediction, and
more than two times larger than the
recent lattice QCD calculations.

\end{abstract}

~~~PACS numbers: 11.55.Hx, 11.30.Hv, 12.38.t, 14.20.Jn
\end{titlepage}

\section{Introduction}

Flavor symmetry plays essential role in classification of the hadrons. The
light hadronic states are successfully described by using SU(3) flavor
symmetry. In the case this symmetry is exact, hadrons belonging to the same
representation of SU(3) flavor group could be degenerate. Experimentally it
is known that the hadrons belonging to the same representation have
different masses, which leads to SU(3) flavor symmetry breaking. At quark
level, this symmetry is broken due to the mass difference of the light u, d
and s quarks.

The breaking of the SU(3) flavor symmetry might lead to mixing of hadrons.
In other words, the definite flavor eigenstates can mix to form the
physically observed states.

Long time ago, it is observed that the lowest lying hadrons $\Lambda$ and
$\Sigma$ can be represented as the combination of the SU(3) octet, pure
isospin $I=0~(\Lambda)$, and  $I=1~(\Sigma^0)$ baryons in the following form
\cite{Rmix01},
\bea
\label{emix01}
\Lambda \es \Lambda^0 \cos\alpha - \Sigma^0 \sin\alpha~, \nnb\\
\Sigma  \es \Lambda^0 \sin\alpha + \Sigma^0 \cos\alpha~.
\eea
The $\Sigma-\Lambda$ mixing angle is estimated in framework of the quark
model whose value is predicted to be $0.86^0$ \cite{Rmix01,Rmix02}.
(see also \cite{Rmix03}). Very recently, the
lattice QCD (LQCD) group presented the latest estimate on $\Sigma-\Lambda$
mixing angle to have the value $\alpha=0.40$ \cite{Rmix04}, which is approximately two
times smaller compared to the prediction of the quark model.

The aim of the present note is to determine the $\Sigma$--$\Lambda$ mixing
angle within the QCD sum rules, and compare this mixing angle with the
predictions of the quark model and LQCD.

In determination of the $\Sigma$--$\Lambda$ mixing angle
within the QCD sum rules we follow the method suggested in \cite{Rmix05}, and
for this goal we
start by considering the following correlation function,
\bea
\label{emix02}
\Pi = i \int d^x e^{ipx} \lla 0 \vel \mbox{T} \left\{\eta_H(x) \bar{\eta}_H(0)
\right\} \ver 0 \rra~,
\eea
where T is the time ordering operator, $\eta_H$ is the interpolating
current, carrying the same quantum numbers as the corresponding hadron. If
the bare $H_1^0$ and $H_2^0$ states are mixed, the corresponding physical states
with definite mass should be the linear combinations of these bare states.
In this case, the interpolating currents corresponding to the physical
states could be represented as the superposition of the interpolating
currents corresponding to the bare states, i.e.,
\bea
\label{emix03}
\eta_\Lambda \es \sin\alpha \eta_{\Lambda^0} + \cos\alpha \eta_{\Sigma^0} ~, \nnb \\
\eta_\Sigma  \es \cos\alpha \eta_{\Lambda^0} - \sin\alpha \eta_{\Sigma^0} ~,
\eea
where $\alpha$ is the mixing angle between $\Lambda^0$ and $\Sigma^0$
states. In presence of only two physical states, Eq. (\ref{emix02}) can be
written as,
\bea
\label{emix04}
\Pi = i \int d^4x e^{ipx} \lla 0 \vel \mbox{T} \left\{\eta_\Lambda (x)
\bar{\eta}_\Sigma \right\} \ver 0 \rra~.
\eea
It should be remembered that the general form of the correlator function is,
\bea
\label{nolabel01}
\Pi(p) = \Pi_1(p^2) \not\!{p} + \Pi_2(p^2) I~,\nnb
\eea
and coefficients of the $\not\!{p}$ and $I$ (unit operator) structures,
i.e., $\Pi_1(p^2)$ and $\Pi_2(p^2)$ can both be used in determining the mixing
angle.

In order to construct the sum rules for the mixing angle $\alpha$, the
correlation function (\ref{emix04}) is calculated in terms of hadrons,
quarks and gluons. Using the duality ansatz these two representations are
matched and the sum rules for the corresponding physical quantity is
obtained.

The hadronic representation of the correlation function is obtained by
saturating it with the full set of baryons having the same quantum numbers
as the corresponding interpolating current. Since $\eta_{H_1}$ and
$\eta_{H_2}$ can create only the states $H_1$ and $H_2$, correspondingly,
the hadronic part of the correlation function is obviously zero. Using Eq.
(\ref{emix03}) in Eq. (\ref{emix04}), one can easily obtain the expression
for the mixing angle for both structures,
\bea
\label{emix05}
\tan 2\alpha = { 2 \Pi_{\Sigma\Lambda}^0 \over 
\Pi_{\Sigma\Sigma}^0 - \Pi_{\Lambda\Lambda}^0}~,
\eea
where $\Pi_{ij}^0$ are the correlation functions corresponding to the
unmixed states, i.e.,
\bea
\label{emix06}
\Pi_{ij}^0 = i \int d^4x e^{ipx} \lla 0 \vel \mbox{T} \left\{ \eta_i^0 (x)
\eta_j^0 \right\} \ver 0 \rra~,
\eea
where $(i,j=\Lambda^0$ or $\Sigma^0)$. So the problem of determination of
the mixing angle requires the calculation of the theoretical part of the
correlation function, for which the expressions of the interpolating
currents are needed.

According to the $SU(3)_f$ classification
the interpolating currents for the unmixed $\Lambda^0$ and $\Sigma^0$ are
chosen as,
\bea
\label{emix07}
\eta_{\Lambda^0} \es 2 \sqrt{1\over 6} \varepsilon^{abc} \Big\{2 (u^{aT} C
d^b) \gamma_5 s^c + 2 \beta (u^{aT} C \gamma_5 d^b) s^c +
(u^{aT} C s^b)\gamma_5 d^c + \beta (u^{aT} C \gamma_5 s^b) d^c \nnb \\
\ek (d^{aT} C s^b)\gamma_5 u^c - \beta (d^{aT} C \gamma_5 s^b) u^c \Big\}~,\nnb \\
\eta_{\Sigma^0} \es  \sqrt{2} \varepsilon^{abc} \Big\{(u^{aT} C s^b) \gamma_5 d^c
+ \beta (u^{aT} C \gamma_5 s^b) d^c + (d^{aT} C s^b) \gamma_5 u^c +
\beta (d^{aT} C \gamma_5 s^b) u^c \Big\}~,
\eea
where $a,b,c$ are the color indices, $C$ is the charge conjugation
operator, and $\beta$ is the arbitrary constant with $\beta=-1$
corresponding to the so--called Ioffe current.

Using the operator product expansion at $p^2 \!\!\ll\! 0$,
one can easily obtain the expressions for the
correlation functions $\Pi_{11}^0$, $\Pi_{22}^0$, and $\Pi_{12}^0$
from Eq. (\ref{emix03}) from the
QCD side for the $\not\!p$ and $I$ structures. The expressions of these
correlation functions are presented in the Appendix.

In order proceed for the numerical calculations we need the values of
the input parameters that are given as: $\qq (1~GeV) = (-0.246_{-19}^{+28}
~MeV^3)$ \cite{Rmix06}, $\qs = 0.8 \qq$, $\gGgG = 0.47~GeV^4$, $m_0^2 =
(0.8\pm 0.2)~GeV^2$ \cite{Rmix07}. For the masses of the light quarks we use their
$\overline{MS}$ values given as: $\overline{m}_u (1~GeV) = 3.11 ~MeV$,
$\overline{m}_d (1~GeV) = 6.48 ~MeV$,
$\overline{m}_s (1~GeV) = 128.25 ~MeV$ \cite{Rmix03}.   

It follows from the expressions of the invariant functions that in order
to determine the $\Sigma$--$\Lambda$ mixing angle three arbitrary
parameters are involved, namely, the continuum threshold $s_0$, the Borel mass parameter
$M^2$, and the parameter $\beta$ (see the expressions
of the interpolating currents); and of course the mixing angle should be
independent of them all. As is well known, the continuum threshold is
related to the energy of the first excited state. The difference
$\sqrt{s_0}-m_{ground}$, where $m_{ground}$ is the mass of the ground state,
is equal to the energy needed to excite the particle to its first energy
state. This difference usually changes in the range between
$0.3$--$0.8~GeV$. It follows from  the analysis of the mass sum rules that
in order to reproduce the experimental values of the masses of the $\Sigma$
and $\Lambda$ baryons, the continuum threshold $s_0$ should lie in the range
$2.5~GeV^2 \le s_0 \le 3.2~GeV^2$ \cite{Rmix08,Rmix09}.
Moreover, the working region of the Borel mass
parameter should be such that, the results for the $\Sigma$--$\Lambda$
mixing angle should exhibit good stability with respect to the variation of
$M^2$ at fixed values of $s_0$. The upper bound of $M^2$ is obtained by
demanding that the higher states and continuum contributions should be less
than 30\% of the total result. The lower bound of $M^2$ is determined from
the condition that the contributions of higher dimensional operators should
be less than the perturbative one. From these conditions the
working region of $M^2$ is determined to be $1.4~GeV^2 \le M^2 \le
2.2~GeV^2$.

In Figs. (1) and (2), we present the dependence of the mixing angle $\alpha$ on
$M^2$ at the value of the continuum threshold $s_0=3.2~GeV^2$ and,
at several fixed values of the auxiliary parameter
$\beta$, for the coefficients of the structures $\not\!{p}$ and $I$,
respectively. We observe from Fig. (1) that in the range $1.4~GeV^2 \le M^2
\le 2.2~GeV^2$ of the Borel parameter, the mixing angle $\alpha$ exhibits
good stability for the values of the auxiliary parameter $\beta=-3;~\pm 1$ for
the structure $\not\!{p}$. As can be traced from Fig. (2), the mixing angle
$\alpha$ seems to be rather stable at all considered values of the auxiliary
parameter $\beta$ for the structure $I$ at the fixed value of the continuum
threshold $s_0=3.2~GeV^2$. 

Our final attempt for determination of the mixing angle is to find the
region of $\beta$ where the mixing angle exhibits insensitivity to its
variation. For this aim we study the dependence of the mixing angle
$\alpha$ on $\cos\theta$ where $\beta=\tan\theta$, at several fixed values
of $M^2$ and at $s_0=3.2~GeV^2$, and presented them in Figs. (3) and (4) for the
coefficients of the structures $\not\!{p}$ and $I$, respectively.
In this respect, the results of our numerical analysis depicted in Figs. (3)
and (4) can be summarized as follows:
\begin{itemize}

\item For the structure $\not\!{p}$, in the above--determined working
regions of $M^2$ and $s_0$, the best stability for the mixing angle is
achieved when $-1 \le \cos\theta \le -0.5$, and the mixing angle is found to
have the value $\alpha=(1.15 \pm 0.05)^0$.

\item For the structure $I$ not only there is no stability region for the mixing
angle, but also the mixing angle changes its sign. Therefore prediction for
the value of the mixing angle from the structure $I$ is not reliable.

\end{itemize}

Therefore we conclude that, the final result for the mixing angle is $\alpha=     
(1.15 \pm 0.05)^0$ which is obtained from the $\not\!{p}$ structure. The
error in determination of the mixing angle can be attributed to the
uncertainties in the value of the continuum threshold $s_0$,
the quark condensates, and the scale parameter $\Lambda$. The
results presented in this work can further be improved by taking
${\cal O}(\alpha_s)$ corrections in to account.

Finally, we compare our result on the $\Sigma$--$\Lambda$ mixing with the
calculations of the quark and lattice QCD models, whose predictions are,
\bea
\label{nolabel02} 
\alpha \es 0.86^0~,~~\cite{Rmix03} \nnb\\
\alpha \es 0.40^0 \pm 0.026^0~,~~\cite{Rmix04} \nnb
\eea
respectively. From these results we observe that, our prediction on the
$\Sigma$--$\Lambda$ mixing angle is very close to the result obtained in the
quark model, and more than two times larger compared to that of the
result obtained in the lattice QCD model. A reliable lattice QCD
determination of the $\Sigma$--$\Lambda$ mixing angle requires an equally
highly accurate reproduction of the octet baryon mass differences, which has
not yet been established.

In conclusion, the mixing angle between the $\Sigma$ and $\Lambda$ baryons
is estimated within the framework of the light cone sum rules method.
A comparison of our result with the predictions of the quark and lattice QCD
models is presented.

\newpage

\section*{Appendix}
\setcounter{equation}{0}
\setcounter{section}{0}


\section*{$\Pi_{\Sigma^0\Lambda^0}^0 (u,d,s)$ for the structure $\not\!{p}$}

%
%
\bea
&&e^{m_{\Sigma^0}^2/2 M^2} e^{m_{\Lambda^0}^2/2 M^2}
\Pi_{\Sigma^0\Lambda^0}^0 (u,d,s) = \nnb \\
&& - {1\over 3072 \sqrt{3} \pi^2 M^4} (1 -  \beta)  
\Big[ (3 m_s + 4  \beta m_s + m_u +   \beta m_u) \dd \nnb \\
\ar (2 + 3 \beta)(m_d - m_u) \sp - (m_d + \beta m_d + 3 m_s +
4 \beta m_s) \uu \Big] \gGgG  m_0^2 \nnb \\
\ar {1\over 1152 \sqrt{3} \pi^2 M^2} \Bigg\{ \Big[ 3(1 +  \beta +  \beta^2) m_d + 
     2(1 -  \beta) \Big(2 +  \beta - 3(1 +  \beta) \gamma_E\Big) m_s \nnb \\
\ek 3(1 -  \beta^2) \gamma_E m_u \Big] \gGgG \dd + 
   (1 -  \beta) \Big[2(2 +  \beta) - 3 (1 +  \beta) \gamma_E\Big] \gGgG (m_d - m_u) \sp \nnb \\
\ek 12 \pi^2 (1 -  \beta) (7 + 5 \beta) m_0^2 \dd \sp - 
   \Big[ (1 -  \beta) \Big( 2 (2 +  \beta) m_s \nnb \\
\ek 3 (1 +  \beta) \gamma_E
        (m_d + 2m_s)\Big) \gGgG  + 3 (1 +  \beta +  \beta^2) \gGgG m_u \nnb \\
\ek 12 \pi^2 (1 -  \beta) (7 + 5 \beta) m_0^2 \sp \Big] \uu + 
   3 (1 -  \beta^2) \gGgG \Big[\dd(2m_s + m_u) \nnb \\
\ar (m_d - m_u) \sp - 
     (m_d + 2 m_s) \uu\Big] \ln {M^2\over \Lambda^2}\Bigg\} \nnb \\
\ar {1\over 128 \pi^2} \sqrt{3} (1 -  \beta^2) \Bigg(\gamma_E - \ln {M^2\over \Lambda^2} \Bigg) m_0^2
  \Big(m_s \dd  + m_d\sp - m_u\sp - m_s\uu\Big) \nnb \\
\ek {1\over 32 \sqrt{3}\pi^2} M^2 (2 -  \beta -  \beta^2) \Big(m_s \dd + m_d\sp - m_u\sp -
m_s\uu\Big) \nnb \\
\ek {1\over 384 \sqrt{3} \pi^2} \Bigg\{ 3 \Big[ 2 (1 +  \beta +  \beta^2) m_d - 6 m_s + m_u + 
      \beta(m_s + 5  \beta m_s -   \beta m_u) \Big] m_0^2 \dd \nnb \\
\ek 32 \pi^2 (2 -  \beta -  \beta^2) \Big(\dd - \uu \Big)\sp \nnb \\
\ek 3 \Big[ (7 -  \beta - 6 \beta^2) (m_d - m_u) \sp + 
     \Big( (1 -  \beta) (m_d +   \beta m_d - 6 m_s - 5  \beta m_s) \nnb \\
\ar 2 (1 +  \beta +  \beta^2) m_u\Big) \uu \Big] m_0^2 \Bigg\}~. \nnb
\eea
\section*{$\Big(\Pi_{\Sigma^0\Sigma^0}^0 - \Pi_{\Lambda^0\Lambda^0}^0 \Big)
(u,d,s)$ for the structure $\not\!{p}$}
\bea
&&e^{m_{\Sigma^0}^2/2 M^2} e^{m_{\Lambda^0}^2/2 M^2}
\Big( \Pi_{\Sigma^0\Sigma^0}^0 - \Pi_{\Lambda^0\Lambda^0}^0 \Big)(u,d,s) = \nnb \\
&&{1\over 4608 \pi^2 M^4} (1 - \beta)  \Big\{  \Big[m_s + 2 \beta m_s -
(5 + 7 \beta) m_u\Big] \dd \nnb \\
\ar (4 + 5 \beta) (m_d + m_u) \sp - \Big[(5 + 7 \beta) m_d - (1 +
2 \beta) m_s\Big] \uu \Big\} \gGgG m_0^2\nnb \\
\ek {1\over 1728 \pi^2 M^2} \Bigg\{ \Big[ (1 - \beta) \Big(2 (2 + \beta) 
- 9 (1 + \beta) \gamma_E \Big) 
       (m_u+m_d)  + 6 (1 + \beta + \beta^2) m_s \Big] \gGgG \sp \nnb \\
\ek \Big\{ \Big[3 (1 + \beta + \beta^2) m_d -
 (1 - \beta) \Big(2 (2 + \beta) m_s - 
         [4 (2 + \beta) - 9 (1 + \beta) \gamma_E] m_u\Big)\Big] \gGgG  \nnb \\
\ar 12 \pi^2 (1 - \beta) (7 + 5 \beta) m_0^2 \Big(\sp - 2 \uu\Big) \Big\} \dd - 
   \Big\{(1 - \beta) \Big[ \Big(4 (2 + \beta) \nnb \\
\ek 9 (1 + \beta) \gamma_E \Big) m_d - 
       2 (2 + \beta) m_s \Big] \gGgG  + 3 (1 + \beta + \beta^2) \gGgG m_u \nnb \\
\ar 12 \pi^2 (1 - \beta) (7 + 5 \beta) m_0^2  \sp\Big\} \uu \nnb \\
\ek 9 (1 - \beta^2)  \Big[ \dd m_u - (m_d + m_u) \sp + m_d \uu \Big] 
\gGgG \ln {M^2\over \Lambda^2} \Bigg\} \nnb \\
\ek {1\over 64 \pi^2} (1 - \beta^2) \Bigg( \gamma_E - \ln {M^2\over \Lambda^2} \Bigg) m_0^2
\Big[\dd (m_s - 2 m_u) + m_u \sp + m_d \Big(\sp - 2 \uu\Big) + m_s \uu \Big] \nnb \\
\ar {1\over 48 \pi^2} M^2 (2 - \beta - \beta^2) \Big [\dd (m_s - 2 m_u) + m_u \sp + 
   m_d \Big(\sp - 2 \uu\Big) + m_s \uu \Big] \nnb \\
\ek {1\over 576 \pi^2} \Bigg\{\Big[ 3 \Big (2 (1 + \beta + \beta^2) m_d + 
       (1 - \beta) (8 m_s + 7 \beta m_s - 13 m_u - 11 \beta m_u)\Big) m_0^2 \nnb \\
\ar 32 \pi^2 (2 - \beta - \beta^2) \Big(\sp - 2 \uu\Big)\Big] \dd + 32 \pi^2 (2 - \beta - \beta^2)
    \sp \uu \nnb \\
\ek  3 \Big[4 (1 + \beta + \beta^2) m_s \sp - (5 - \beta - 4 \beta^2) m_u 
      \sp - (8 - \beta - 7 \beta^2) m_s \uu \nnb \\
\ek 2 (1 + \beta + \beta^2) m_u \uu - 
     (1 - \beta) m_d \Big( (5 + 4 \beta) \sp - (13 + 11 \beta) \uu\Big) \Big] m_0^2
\Bigg\}~. \nnb
\eea
\section*{$\Pi_{\Sigma^0\Lambda^0}^0 (u,d,s)$ for the structure $I$}
\bea
&&e^{m_{\Sigma^0}^2/2 M^2} e^{m_{\Lambda^0}^2/2 M^2}
\Pi_{\Sigma^0\Lambda^0}^0 (u,d,s) = \nnb \\
\ek {1\over 128 \sqrt{3} \pi^4} M^6 (2 - \beta - \beta^2) (m_d - m_u) \nnb \\
\ar {1\over 32 \sqrt{3} \pi^2} M^4 (2 - \beta - \beta^2) \Big(\dd - \uu\Big) \nnb \\
\ek {1\over 512 \sqrt{3} \pi^4} M^2 (1 - \beta) (1 + 2 \beta) 
\Bigg( \gamma_E - \ln {M^2\over \Lambda^2} \Bigg) (m_d - m_u) \gGgG \nnb \\
\ek {1\over 768 \sqrt{3} \pi^4} M^2 (1 - \beta) \Big[(1 + 5 \beta) (m_d - m_u) \gGgG + 
   18 \pi^2 (1 + \beta) \Big(\dd - \uu\Big) m_0^2 \Big] \nnb \\
\ar {1\over 12288 \sqrt{3} \pi^4 M^2} \Big[(1 - \beta) (1 + 2 \beta)
(m_d - m_u) \gGgG^2 \nnb \\
\ar 512 \pi^4 (1 + \beta + \beta^2) \Big(\dd m_u - m_d \uu\Big) m_0^2 \sp \Big] \nnb \\
\ek {1\over 384 \sqrt{3} \pi^2} (1 - \beta) \Big\{ \Big[(1 + 2 \beta) \gGgG +
16 \pi^2 (2 + \beta) m_s \sp \Big] \uu \nnb \\
\ek  \Big[ (1 + 2 \beta) \gGgG + 16 \pi^2 (2 + \beta) \Big(m_s \sp - (m_d - m_u)
\uu \Big) \dd \Big\}~. \nnb
\eea
\section*{$\Big(\Pi_{\Sigma^0\Sigma^0}^0 - \Pi_{\Lambda^0\Lambda^0}^0 \Big)
(u,d,s)$ for the structure $I$}
\bea
&&e^{m_{\Sigma^0}^2/2 M^2} e^{m_{\Lambda^0}^2/2 M^2}
\Big( \Pi_{\Sigma^0\Sigma^0}^0 - \Pi_{\Lambda^0\Lambda^0}^0 \Big)(u,d,s) = \nnb \\
\ek {1\over 192 \pi^4} M^6 (2 - \beta - \beta^2) (m_d - 2 m_s + m_u) \nnb \\
\ar {1\over 48 \pi^2} M^4 (2 - \beta - \beta^2) \Big(\dd - 2 \sp + \uu\Big) \nnb \\
\ek {1\over 768 \pi^4} M^2 (1 - \beta) (1 + 2 \beta) \Bigg( \gamma_E -
\ln {M^2\over \Lambda^2} \Bigg) (m_d - 2 m_s + m_u) \gGgG \nnb \\
\ek {1\over 1152 \pi^4} M^2 (1 - \beta) \Big[(1 + 5 \beta) (m_d - 2 m_s + m_u) \gGgG \nnb \\ 
\ar 18 \pi^2 (1 + \beta) \Big(\dd - 2 \sp + \uu\Big) m_0^2 \Big] \nnb \\
\ar {1\over 18432 \pi^4 M^2} \Big\{ (1 - \beta) (1 + 2 \beta) (m_d - 2 m_s + m_u)
\gGgG^2 \nnb \\
\ek 512 \pi^4 (1 + \beta + \beta^2) \Big(m_u \dd \sp - 2 m_s \dd \uu + m_d \sp
\uu\Big) m_0^2\Big\} \nnb \\
\ar {1\over 576\pi^2 } (1 - \beta) \Big\{16 \pi^2 (2 + \beta) (m_s - 2 m_u) \sp \uu -
  (1 + 2 \beta) \Big(2\sp - \uu\Big) \gGgG \nnb \\
\ar \Big[(1 + 2 \beta) \gGgG - 16 \pi^2 (2 + \beta) \Big((2 m_d - m_s) \sp - 
       (m_d + m_u) \uu \Big) \Big] \dd \Big\}~, \nnb
\eea
where $M^2$ is the Borel parameter and $\Lambda$ is the energy cut off
separating perturbative and nonperturbative regimes; and $\gamma_E$ is the
Euler constant.

Note that the scale parameter $\Lambda$ is calculated in \cite{Rmix10} whose
value is in the range $0.5 \div 1.0~GeV$.   

\newpage

\section*{Figure captions}
{\bf Fig. 1} Dependence of the $\Lambda$--$\Sigma$ mixing angle
on the Borel mass parameter $M^2$ at the fixed value of the continuum
threshold $s_0=3.2~GeV^2$, and at several fixed values of the auxiliary
parameter $\beta$, for the structure $\not\!{p}$. \\ \\
{\bf Fig. 2} The same as in Fig. 1, but for the structure $I$.\\ \\
{\bf Fig. 3} Dependence of the $\Lambda$--$\Sigma$ mixing angle
on $\cos\theta$ at the fixed value of the continuum       
threshold $s_0=3.2~GeV^2$, and at several fixed values of
the Borel mass parameter $M^2$, for the structure $\not\!{p}$. \\ \\ 
{\bf Fig. 4} The same as in Fig. 1, but for the
structure $I$.

\newpage

\begin{figure}  
\vskip 3. cm
    \includegraphics{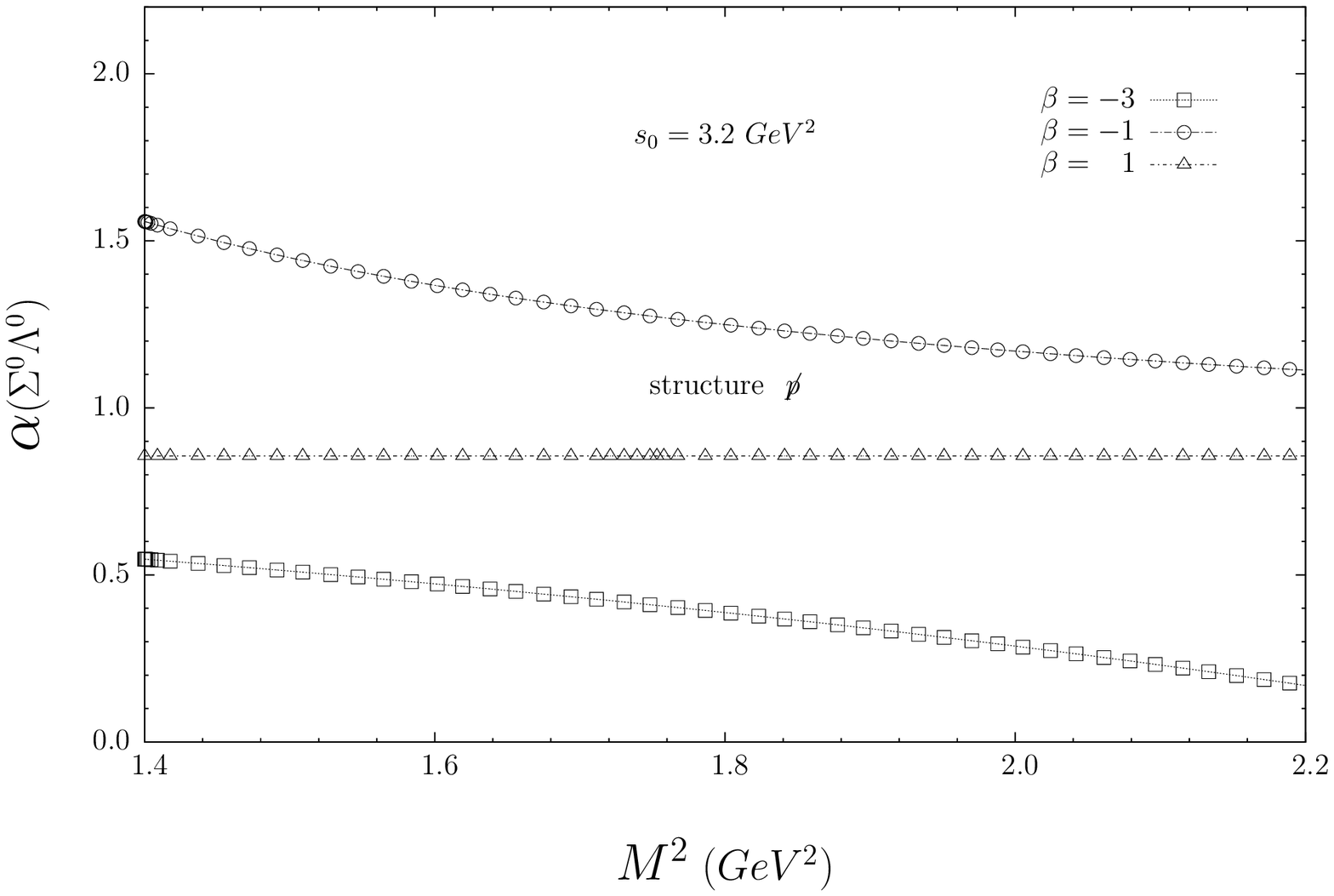}
\vskip 7.0cm   
\caption{}
\end{figure}

\begin{figure}
\vskip 3. cm
    \includegraphics{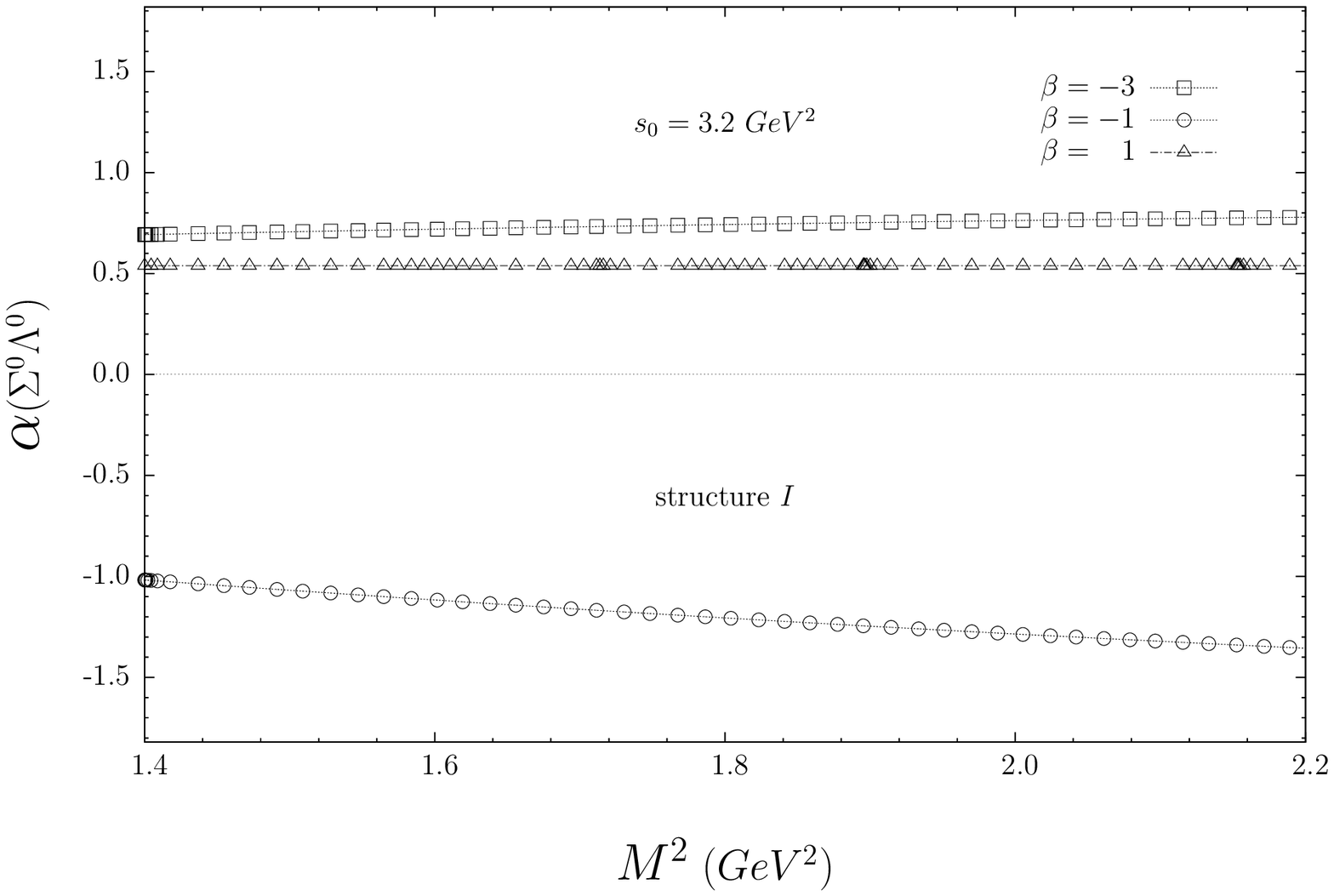}
\vskip 7.0cm
\caption{}
\end{figure}

\begin{figure}  
\vskip 3. cm
    \includegraphics{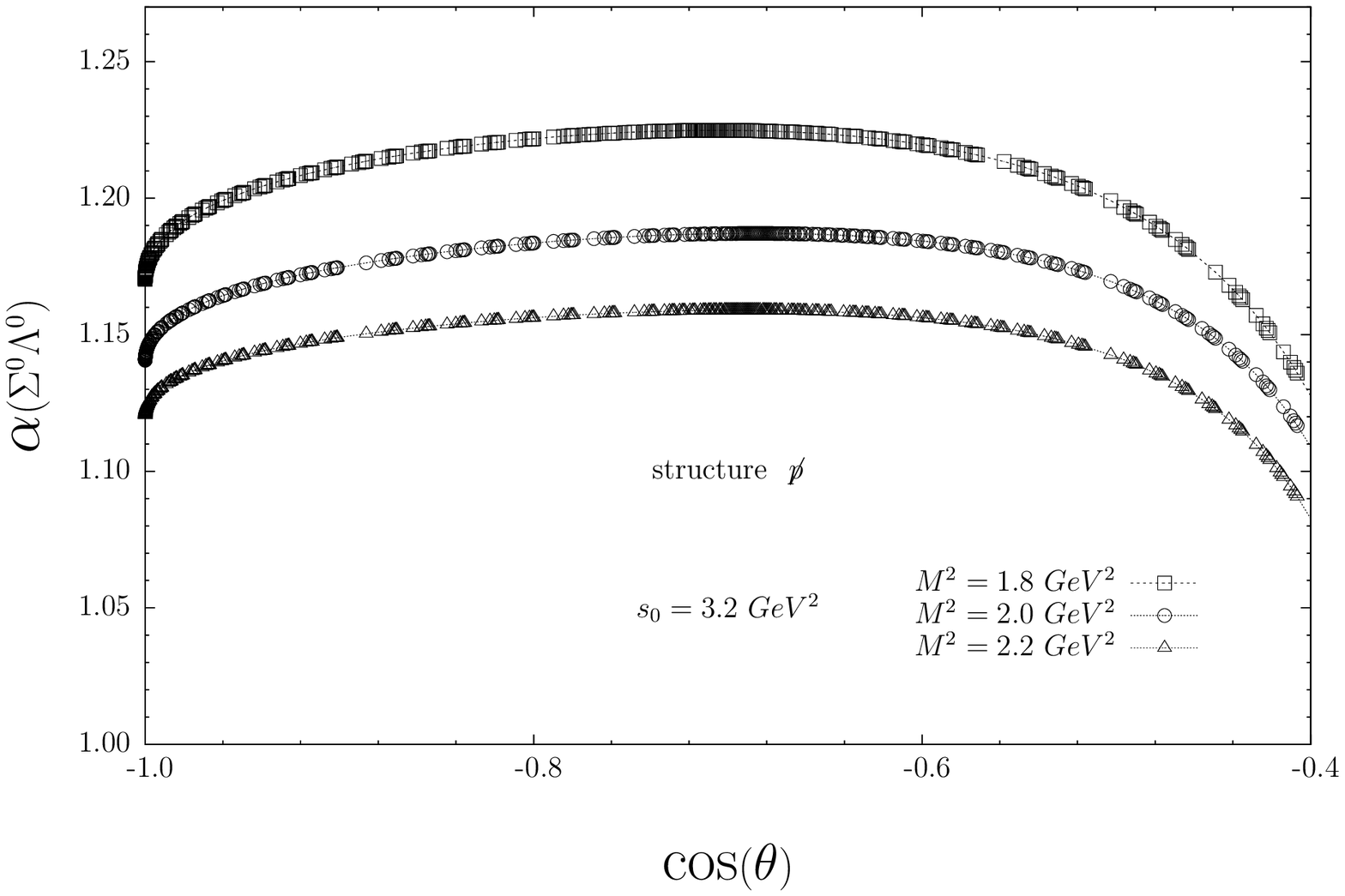}
\vskip 7.0cm   
\caption{}
\end{figure}

\begin{figure}
\vskip 3. cm
    \includegraphics{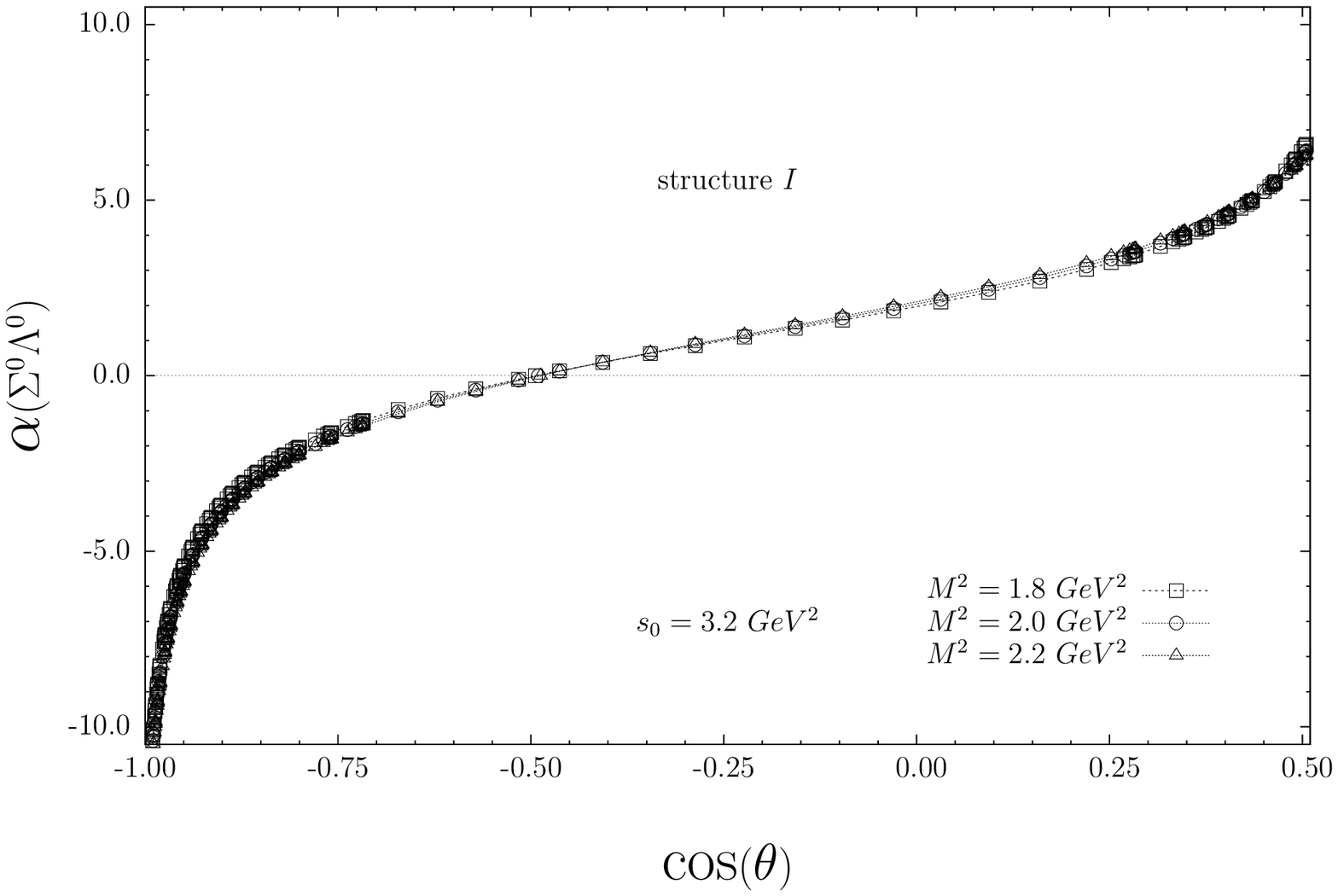}
\vskip 7.0cm
\caption{}
\end{figure}

\end{document}